\documentclass[%
reprint,
superscriptaddress,
preprintnumbers,
amssymb,
aps,
prb,
longbibliography,
]
{revtex4-1}
\usepackage[english]{babel} %francais, polish, spanish, ...
\usepackage [table]{xcolor}
\usepackage{graphicx} %%For loading graphic files
\usepackage{amsmath}
\usepackage{amsthm}
\usepackage{amsfonts}
\usepackage{color}
\usepackage{natbib}

\usepackage{float}
\usepackage{textcomp}
\usepackage{graphicx}% Include figure files
\usepackage{dcolumn}% Align table columns on decimal point
\usepackage{bm}% bold math
\usepackage[mathlines]{lineno}% Enable numbering of text and display math
\usepackage{array}

\begin{document}

%% Title Page %%%%%%%%%%%%%%%%%%%%%%%%%%%%%%%%%%%%%%%%%%%%%%%
%% ==> Write your text here or include other files.

%Title of paper
%\preprint{APS/123-QED}

\title{Low-loss superconducting nanowire circuits using a neon focused ion beam} %Low-loss superconducting nanowires within superconducting resonators using a neon focused ion beam}

\author{J.~Burnett}
\email{burnett@chalmers.se}
\author{J.~Sagar}
\author{O.~W.~Kennedy}
\author{P.~A.~Warburton}
\author{J.~C.~Fenton}
\email{j.fenton@ucl.ac.uk}

\affiliation{London Centre for Nanotechnology, UCL, 17-19 Gordon Street, London, WC1H 0AH, UK}

\selectlanguage{english}

\date{\today}

\begin{abstract}
We present low-temperature measurements of low-loss superconducting nanowire-embedded resonators in the low-power limit relevant for quantum circuits. The superconducting resonators are embedded with superconducting nanowires with widths down to 20~nm using a neon focused ion beam. In the low-power limit, we demonstrate an internal quality factor up to 3.9$\times$10$^5$ at 300~mK (implying a TLS-limited quality factor up to 2$\times$10$^5$ at 10~mK), not only significantly higher than in similar devices, but also matching the state of the art of conventional Josephson-junction-embedded resonators. %The power- and temperature-dependence of the losses are well described by existing theories of superconducting devices in the presence of two-level systems, showing that the neon focused ion beam induces negligible additional loss. 
We also show a high sensitivity of the nanowire to stray infrared photons, which is controllable by suitable precautions to minimise stray photons in the sample environment. Our results suggest that there are excellent prospects for superconducting-nanowire-based quantum circuits.
\end{abstract}

\maketitle

\section{Introduction}

Quantum circuits based on conventional Josephson-junctions have begun to tackle real-world problems\cite{PhysRevX.6.031007}. This has been despite high decoherence produced by the loss\cite{lindstrom2009properties,macha2010losses} and noise\cite{burnett2014evidence,ramanayaka2015evidence}  caused by parasitic two-level systems (TLS)\cite{phillips1972tunneling,faoro2015interacting}. In principle, superconducting nanowires can provide a route to low-decoherence quantum circuits due to their monolithic structure and lack of a TLS-hosting oxide layer. To date, superconducting nanowires with cross-sectional areas approaching the coherence length have demonstrated a variety of Josephson\cite{hao2009characteristics,levenson2011nonlinear} and phase-slip\cite{astafiev2012coherent,peltonen2016coherent,belkin2011little,webster2013nbsi} effects, but features such as their unconventional current-phase relationships\cite{likharev1979superconducting} remain unexploited in quantum circuits. Previous demonstrations of superconducting nanowire-embedded resonators exhibit unusually high dissipation, with internal quality factors ($Q_{\rm i}$) below 5$\times$10$^3$,\cite{belkin2011little,astafiev2012coherent,jenkins2014nanoscale,peltonen2016coherent}, far lower than in similar conventional Josephson-junction-based circuits\cite{de2013charge,simoen2015characterization}. In general, the performance of nanowire-embedded resonators can be limited by material quality, interface imperfections, resist residues and the measurement environment.

We demonstrate superconducting nanowire-embedded circuits with single photon $Q_{\rm i}$ up to 3.9$\times$10$^5$, comparable to or even better than conventional Josephson-junction resonators. Superconducting nanowires with widths down to 20~nm were fabricated with a neon focused ion beam (FIB). We study the loss in our devices within the well-established framework of loss mechanisms in superconducting resonators\cite{lindstrom2009properties,macha2010losses,faoro2015interacting,quintana2014characterization,goetz2016loss} to determine which factors are significant in limiting their performance. The vastly improved $Q_{\rm i}$ demonstrates that the detrimental effects can be sufficiently reduced and shows that competitive quantum circuits could be based on monolithic nanowire technology. %Suggesting that excess dissipation in previous devices was linked to the nanowire fabrication\cite{astafiev2012coherent,peltonen2016coherent,belkin2011little,jenkins2014nanoscale}%, possibly due to TLS from resist residues\cite{astafiev2012coherent,peltonen2016coherent}, TLS in substrate surface oxides\cite{belkin2011little} or damage from a gallium FIB\cite{jenkins2014nanoscale}.

\section{Methods}

Superconducting 20-nm-thick NbN films were deposited on sapphire by dc magnetron sputtering from a 99.99\%-pure Nb target in a 1:1 Ar:N$_2$ atmosphere. The vacuum chamber was pumped to 6$\times$10$^{-7}$~mbar before sputtering at a pressure of 3.5$\times$10$^{-3}$~mbar and power of 200~W. The superconducting critical temperature, $T_{\rm c}$, was 10~K with a sheet resistance of 450~$\Omega/sq$. Electron-beam lithography (EBL) was used to pattern $\lambda/4$ and $\lambda/2$ coplanar microwave resonators capacitively coupled to a common microwave feed line (shown in Fig.~\ref{fig1}d). The width of the central conductor was 10~$\mu$m and the gap was 5~$\mu$m. This pattern was transferred from a 300-nm-thick-layer of polymethyl methacrylate (PMMA) into the film by a reactive ion etch (RIE) using a 2:1 ratio of SF$_{6}$:Ar, at 30~W and 30~mbar. %The low power and pressure improve the etch selectivity to the PMMA.

A neon FIB was used to directly pattern\cite{timilsina2014monte} nanowires in the central conductor of the microwave resonators at the current antinode -~see Fig.~\ref{fig1}b. With an acceleration voltage of 15~kV, the clearance dose for the NbN film is $\approx$~0.3~nC/$\mu$m$^2$. 15~kV was chosen as a compromise between minimising the spot size and minimising lateral milling of the nanowire\cite{rahman2012prospects}, leading to a few-minute mill time per $\mu$m$^2$ for a $\sim$1~pA  beam current. By prior patterning of a sub-200-nm-wide precursor wire in the same EBL step as the resonator (shown in Fig.~\ref{fig1}c), we minimise the mill time and the total neon flux that the nanowire is subject to. Several devices were measured, and Table~\ref{ResTab1} shows important parameters including nanowire dimensions. The nanowire devices all feature two nanowires, configured either in parallel so that the nanowires complete a superconducting loop\cite{belkin2011little}, or in series with a wider segment in between\cite{hongisto2012single}. Here, there is no external flux- or gate-bias, so the nanowires are treated as simple constrictions within the superconductor. 

%15kV was chosen as a suitable accelerating voltage for the Ne+ ions to minimise lateral milling\cite{rahman2012prospects} of the NbN structures whilst maintaining a small probe size. 

Samples were enclosed within a brass box and cooled using a $^3$He refrigerator containing a heavily attenuated microwave in-line and an out-line with a cryogenic high-electron-mobility transistor (HEMT) amplifier. 

\section {Results \& Discussion}

\begin{figure*}
	\includegraphics[width=2\columnwidth]{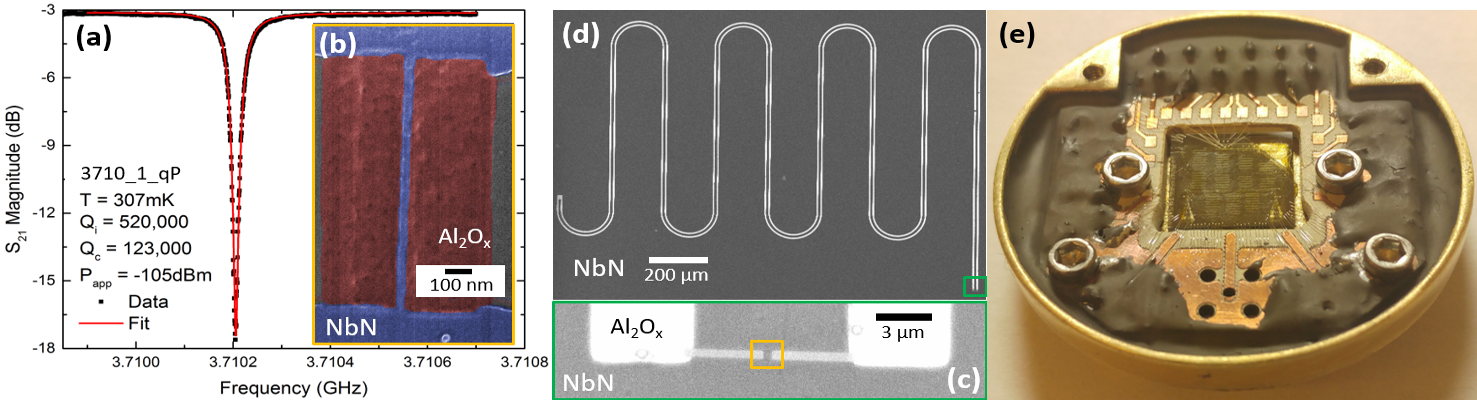}
	\caption{({\bf a}) The S$_{21}$ magnitude response of the nanowire-embedded resonator 3710\_1qP. The red line is a fit to Eq.~\ref{reseq}. ({\bf b}) A false-colour He FIB micrograph of a neon FIB milled nanowire (3710\_1qP) with dimensions of 27~nm by 1.2~$\mu$m. The NbN is shown in blue, while the milled region is shown in red. ({\bf c, d}) Scanning electron micrographs of ({\bf c}) the shorted end of a $\lambda/4$ resonator before milling by neon FIB. ({\bf d}) the whole $\lambda/4$ resonator. ({\bf e}) A photograph of the sample holder. In the centre is a chip, which is wirebonded to a microwave printed circuit board, the dark material is Eccosorb.}
	\label{fig1}
\end{figure*}

Figure~\ref{fig1}a shows the forward transmission ($S_{21}$) magnitude response of a nanowire-embedded resonator, at 307~mK and for an applied microwave drive of $-105$~dBm, demonstrating $Q_{\rm i}$~=~5.2$\times$10$^5$. This $Q_{\rm i}$ is significantly higher than in comparable nanowire-based devices\cite{belkin2011little,astafiev2012coherent,jenkins2014nanoscale,peltonen2016coherent}. This highlights the promise of the neon FIB and demonstrates that superconducting nanowires are not intrinsically lossy. %indicates that the neon FIB is a promising approach for realising low-loss superconducting nanowires and % and that the dominant loss mechanism of the previously studied nanowires was possibly introduced through the fabrication. 
The complex $S_{21}$ notch response of the superconducting resonators is fitted by\cite{probst2015efficient}
\begin{equation}S_{21}^{\rm}{(\nu)} = ae^{j\theta}e^{-2\pi{j}\nu\tau}\left[1-\frac{(Q_{\rm L}/|Q_{\rm c}|)e^{j\phi}}{1+2jQ_{\rm L}(\nu/\nu_{0}-1)}\right],
\label{reseq}
\end{equation}
where $\nu$ is the applied frequency, $\nu_{0}$ the resonance frequency, $Q_{\rm L}$ the loaded quality factor and $|Q_{\rm c}|$ the absolute value of the coupling quality factor; $\phi$ accounts for impedance mismatches, $a$ describes a change in amplitude, $\theta$ describes a change in phase and $\tau$ a change in the electronic delay. %The handling of free parameters in Eq.~\ref{reseq} is described in Ref.~\citenum{probst2015efficient}. 
The internal quality factor, $Q_{\rm i}$, is defined by $1/Q_{\rm L} = 1/Q_{\rm i} + \text{\rm Re}(1/Q_{\rm c})$ and the energy within the resonator is $W_{\rm sto} = 2P_{\rm app}S_{\rm min}Q_{\rm L}/{\omega_{0}}$, where $P_{\rm app}$ is the applied microwave power (in W) and $S_{\rm min}$ the normalized minimum of the resonator magnitude response. We describe the microwave power in terms of the average number of photons in the resonator, $\left<n\right>$,  given by $\left<n\right> = W_{\rm sto}/(h\nu_0)$, where $h$ is Planck's constant.

\begin{figure}
	\includegraphics[width=1\columnwidth]{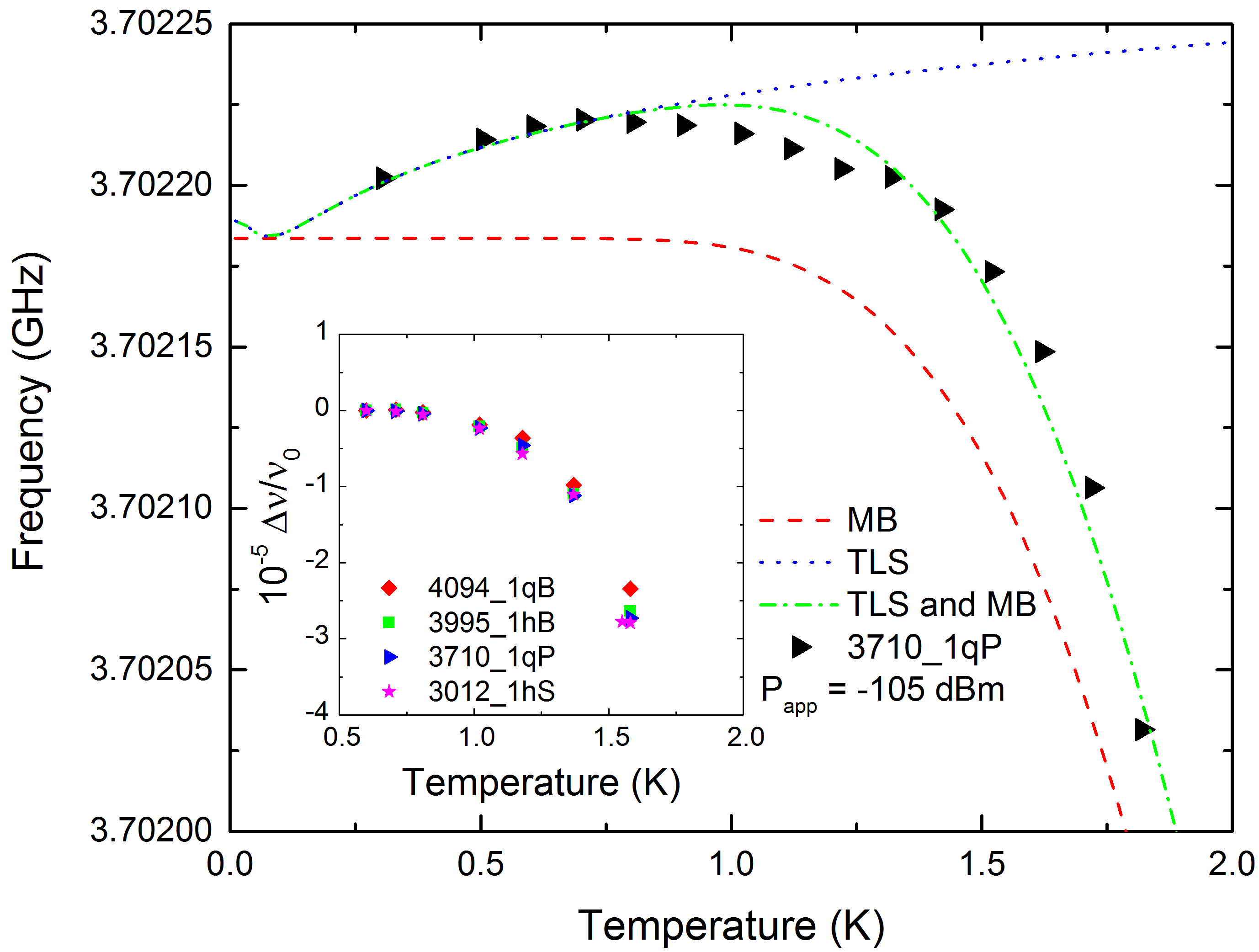}
	\caption{Resonant frequency of resonator 3710\_1\_qP as a function of temperature. Red broken line: Variation arising from kinetic inductance changes described by MB theory. Blue dotted line: Variation arising from TLS losses. Green dash-dotted line: Fit to data including both MB and TLS effects. Inset: Normalised frequency shift as a function of temperature for all resonators on chip 1.}
	\label{fig4}
\end{figure}

To examine the effect of the neon FIB on the NbN film, we measured the resonator response as a function of temperature (shown in Fig.~\ref{fig4}). As temperature decreases from 2~K to 1~K, the resonant frequency increases due to changes in the complex conductivity which are described by $\frac{\Delta\nu}{\nu_{0}} = \frac{\alpha}{2}\frac{\Delta\sigma_{2}}{\sigma_{2}}$, where $\frac{\Delta\nu}{\nu_{0}}$ is the normalised resonance frequency, $\alpha$ is the kinetic inductance fraction and $\sigma_{2}$ is the imaginary part of the complex conductivity as given by Mattis-Bardeen (MB) theory\cite{mattis1958theory}. The inset of Fig.~\ref{fig4} shows the temperature dependence of the resonant frequency for all resonators on chip 1. The bunching of data points indicates a very similar $T_{\rm c}$ whether the resonator contains nanowires or not, implying that the neon FIB has not significantly suppressed the superconductivity.

Further decreasing temperature from 1~K, the resonant frequency decreases due to a thermal desaturation of TLS, which can be described by
\begin{equation}\frac{\Delta\nu}{\nu_{0}} = \frac{F\delta_{\rm TLS}^{\rm i}}{\pi}\left[{\rm Re}\Psi\left(\frac{1}{2}+\frac{1}{2\pi{j}}\frac{h\nu_{0}(T)}{k_B{T}}\right)-{\rm ln}\left(\frac{1}{2}\frac{h\nu_{0}(T_{0})}{k_B{T}}\right)\right],
\label{TLSeq2}
\end{equation}
where $F$ is the filling factor which typically relates to device geometry and electric field density, $T_0$ is a reference temperature, $\Psi$ is the complex digamma function and $F\delta_{\rm TLS}^{\rm i}$ is the intrinsic loss tangent. Fig.~\ref{fig4} shows a fit to both the MB and TLS frequency shifts, and the extracted $F\delta_{\rm TLS}^{\rm i}$ is shown in Table~\ref{ResTab1}. Barends {\it et al}.\cite{barends2009noise} have previously showed that, to determine $F\delta_{\rm TLS}^{\rm i}$ using both MB and TLS models, it is not necessary to obtain data in the temperature range covering the frequency upturn below 100mK seen in the TLS fit curve in Fig.~\ref{fig4}. 

The thermal desaturation of TLS below 1~K results in absorption of microwave photons, leading to a power- and temperature-dependent resonator loss rate\cite{macha2010losses,lindstrom2009properties}. At low microwave drive, the unsaturated TLS dominate the loss, but as the microwave drive increases these TLS become saturated and therefore their loss rate decreases. At high microwave drives, where the TLS are saturated, the loss becomes dominated by residual quasiparticles, with a loss rate $\delta_{\rm qp}$ which is temperature-dependent but assumed to be independent of microwave power\cite{goetz2016loss}. The TLS and quasiparticle loss behaviour can be described by
 \begin{equation}\frac{1}{Q_{\rm i}} = \delta^{\rm i}_{\rm tot} = F\delta^{0}_{\rm TLS}\frac{{\tanh}\left(\frac{h\nu_0}{2k_{\rm B}T}\right)}{\left(1+\left(\frac{\left<n\right>}{n_{\rm c}}\right)\right)^{\beta}} + \delta_{\rm qp},
\label{TLSeq}
\end{equation}
where $n_{\rm c}$ is the number of photons equivalent to the saturation field of the TLS, $\beta$ describes how quickly the TLS saturate with power and $F\delta_{\rm TLS}^{0}$ is the TLS loss tangent ($F\delta_{\rm TLS}^{0}$ is power- and temperature-independent). TLS models were originally based on the anomalous properties of glasses at low temperatures\cite{phillips1972tunneling} and assumed non-interacting TLS, which leads to a prediction of  $\beta$~=~0.5. However, as superconducting circuits have improved, this model has failed to accurately describe the power dependence of dielectric losses: a weaker power dependence with $\beta <$~0.5 is frequently found\cite{macha2010losses,burnett2016analysis,wisbey2010effect,khalil2011loss}. This showed the need to consider TLS interactions \cite{faoro2012internal,faoro2015interacting,burnett2014evidence,burnett2016analysis,ramanayaka2015evidence}, changing the loss model to\cite{faoro2012internal,faoro2015interacting} 
\begin{equation}
	\frac{1}{Q_{i}} = FP_{\gamma}\chi{}\ln\left(\frac{Cn_{c}}{n} + \delta_{\rm qp}'\right)\tanh\left(\frac{h\nu_{0}}{2k_{B}T}\right),
	\label{TLSlog}
	\end{equation}
	where $\chi$ is the dimensionless TLS parameter, $P_{\gamma}$ is the TLS switching rate ratio, $C$ is a large constant and $\delta_{\rm qp}'$ is the log-scaled quasiparticle loss rate.%\cite{qpnote} ($\delta_{\rm qp} = FP_{\gamma}\chi{}\ln(\delta_{\rm qp}')$).}

\begin{figure*}[t]
	\includegraphics[width=2\columnwidth]{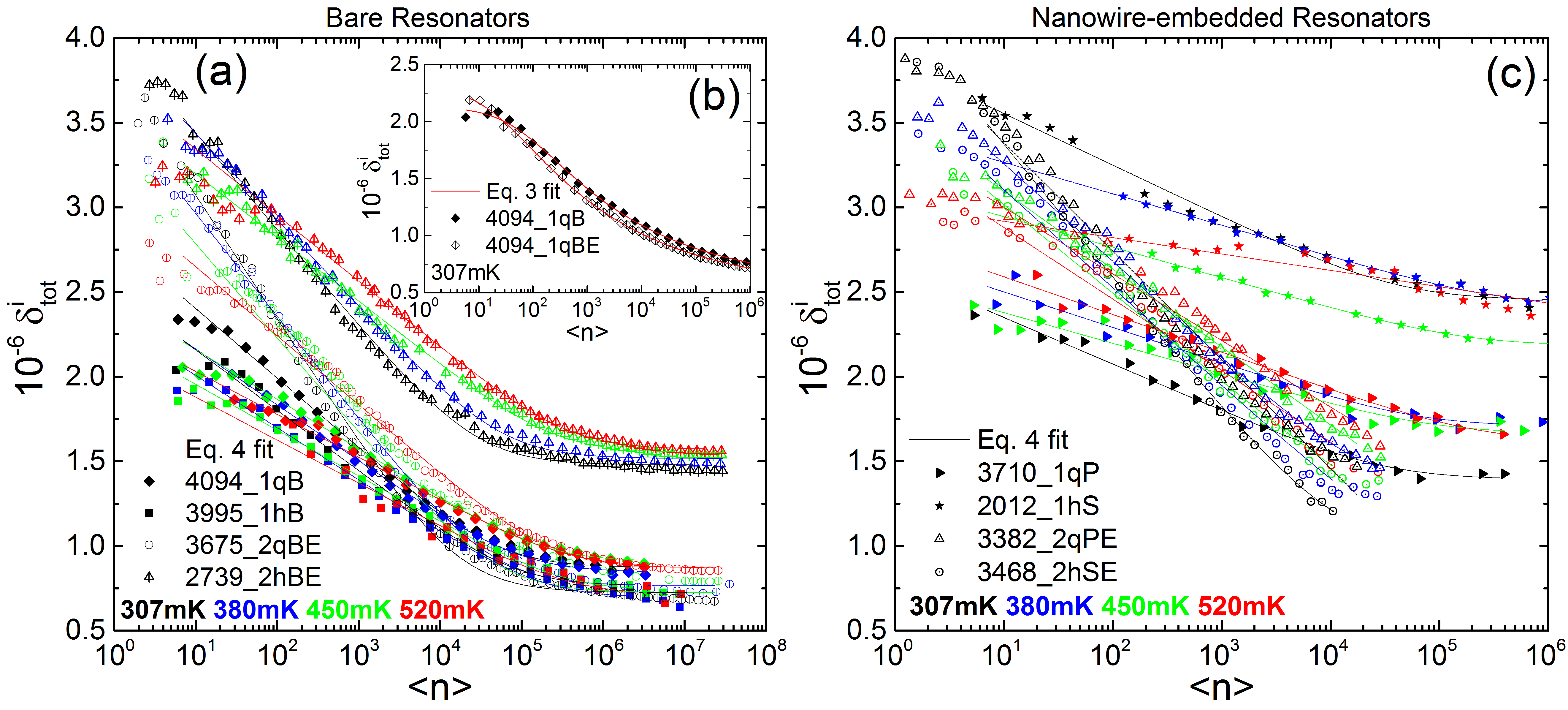}
	\caption{({\bf a}) Loss tangent $\left(\delta^{\rm i}_{\rm tot} = 1/Q_{\rm i}\right)$ as a function of microwave drive for bare resonators. The colours correspond to different temperatures and the hollow symbols indicate the use of an Eccosorb CR-117 lined sample box. The solid lines in all plots represent fits to Eq.~\ref{TLSlog}. ({\bf b}) Loss tangent for the resonator 4094\_1qB with and without the Eccosorb-lined box. ({\bf c}) Loss tangent as a function of microwave drive for the nanowire-embedded resonators.}
	\label{fig3}
\end{figure*}

%To investigate the loss of our devices in more detail, we measured the microwave transmission of the resonator as a function of microwave drive. %At high microwave drive, the resonator's line shape gives information on the resonator's superconducting properties and specifically its critical current.
This loss is examined in more detail by fitting the resonator $S_{21}$ response as a function of microwave drive and temperature (shown in Figs.~\ref{fig3}a--c). Fig.~\ref{fig3}a (Fig.~\ref{fig3}c) show measurements of $\delta^{\rm i}_{\rm tot}$ $\left(\text{{where }}\delta^{\rm i}_{\rm tot} = 1/Q_{\rm i}\right)$  as a function of $\left<n\right>$ on bare (nanowire-embedded) resonators. Each resonator has its own symbol, with solid (hollow) symbols corresponding to measurements in a normal (Eccosorb-lined) sample box. Eccosorb CR-117 (see supplemental\cite{suppnotePRApp}) is a microwave absorber which has been shown to reduce quasiparticle excitation from stray infrared (IR) photons\cite{barends2011minimizing}, the Eccosorb lining is shown in Fig.~\ref{fig1}e.  Different colours correspond to different temperatures. When analysing Figs.~\ref{fig3}a \& c with Eq.~\ref{TLSeq}, we find $\beta \approx$~0.1--0.2 (see supplemental\cite{suppnotePRApp}) implying interacting TLS. The solid lines represent fits to the interacting-TLS model, Eq.~\ref{TLSlog}. Table~\ref{ResTab1} collects fit parameters from both models.

\begin{table}
	\caption{Table of resonator parameters. Resonators are named by $\nu_0$ (MHz), their chip number,  $\lambda/4$ (q) or $\lambda/2$ (h) and whether they are bare resonators (B), have nanowires in series (S), have nanowires in parallel (P) or were measured in an Eccosorb-lined box (E). $\bar{w}$ refers to the nanowire widths. $\delta^{\rm i}_{\rm TLS}$ comes from fits to Eq.~\ref{TLSeq2}, while $\delta^{0}_{\rm TLS}$ \& $\delta_{\rm qp}$ come from fits to Eq.~\ref{TLSeq} and $FP_{\gamma}\chi$ come from fits to Eq.~\ref{TLSlog}.}
	\begin{tabular}{m{1.3cm} m{0.9cm} m{0.9cm} m{0.9cm} m{0.9cm} m{0.9cm} m{0.9cm}}
		\hline	
		Resonator & $\bar{w}$ (nm) & $F\delta^{\rm i}_{\rm TLS}$ ($\times$10$^{-6}$)& $FP_{\gamma}\chi$ ($\times$10$^{-6}$) &$F\delta^{0}_{\rm TLS}$ ($\times$10$^{-6}$)& $\delta_{\rm qp}$ ($\times$10$^{-7}$) \\
		\hline  
		4094\_1qB & - & 6 & 0.57 & 6.6 & 5.6 \\
		3995\_1hB & - & 6.3 & 0.61 & 6.2 & 6.9 \\
		3675\_2qBE & - & 9.8 & 1.11 & 9.5 & 5.6 \\
		2739\_2hBE & - & 12.6 & 1.21 & 10.2 & 13.3 \\
		3710\_1qP & 27, 30 & 4.8 & 0.47 & 5.9 & 12.4 \\
		3012\_1hS & 47, 48 & 6.9 & 0.42 & 7.8 & 21.1 \\
		3382\_2qPE & 20, 23 & 12.9 & 1.13 & 14.3 & 5.4 \\
		3468\_2hSE & 37, 34  & 13.0 & 1.27 & 14.2 & 6.8 \\
	\end{tabular}
	\label{ResTab1}
\end{table}

 %The details of the loss may be described by %a general model of power-independent loss due to excess quasiparticles and power-dependent loss due to TLS (note that $\left<n\right>$ scales with electric field $\mathcal{E}$ as $\left<n\right>\propto\mathcal{E}^2$):

%\begin{figure}
%	\includegraphics[width=1\columnwidth]{fig3}
%	\caption{({\bf a}) and ({\bf b}) Normalised S$_{21}$ magnitude response for the device 3675\_2qBE at 307~mK as a function of microwave drive. ({\bf a}) Standard transmission of the resonator in blue, the onset of nonlinear behaviour in green and the onset of bifurcation in red. ({\bf b}) Transmission across the full range of microwave drives as a contour plot with the resonant notch represented by the dark tones. ({\bf c}) and ({\bf d}) show the same measurement for the nanowire-embedded device 3468\_2hSE at 307~mK. Note that the nonlinear behaviour associated with approaching the critical current appears for $P_{\rm app}$ 30~dB lower for the nanowire-embedded resonator}
%	\label{Sfig2}
%\end{figure}

We first consider bare resonators measured in a standard sample box (solid symbols in Fig.~\ref{fig3}b). Resonators on the same chip show a fabrication-based variability, also found in the literature\cite{bruno2015reducing,goetz2016loss,lindstrom2009properties}: high-$\left<n\right>$ $Q_{\rm i}$~=~1.2--3.1$\times$10$^6$ and low-$\left<n\right>$ $Q_{\rm i}$~=~3.6--5.7$\times$10$^5$ at 307~mK. Increasing the temperature leads to an increase in low-$\left<n\right>$ $Q_{\rm i}$ because, as thermal occupation of the TLS increases, their ability to absorb microwave photons decreases\cite{bruno2015reducing,macha2010losses}, as described by the $\tanh$ temperature term. Increasing temperature also leads to a decrease in high-$\left<n\right>$ $Q_{\rm i}$. This is due to a higher quasiparticle density, meaning that more energy is lost to the quasiparticle system\cite{goetz2016loss}. 

We next consider nanowire-embedded resonators in the standard sample box (solid symbols in Fig.~\ref{fig3}c). At 307~mK, at low $\left<n\right>$, we find $Q_{\rm i}$~=~2.7--3.9$\times$10$^5$, in good agreement with the results from the bare resonators, so the FIB-based fabrication of the nanowire has produced very little additional TLS loss. At high $\left<n\right>$, we find  $Q_{\rm i}$~=~4.1--7.2$\times$10$^5$, 3--5 times lower than the bare resonators, indicating a higher residual quasiparticle density for the nanowire-embedded resonators.%, either due to a suppression of the superconducting properties in the nanowire or because of an increased sensitivity to pair-breaking photons. %To explore this in more detail, we show in Fig.~\ref{fig3}d the temperature dependence of $Q_{\rm i}$ at high $\left<n\right>$. For the normal sample box, $Q_{\rm i}$ does not scale with temperature, indicating that losses are not due to thermally generated quasiparticles.  

Quasiparticles generated from pair-breaking events are an important consideration in conventional Josephson-junction devices\cite{EccosorbCorcoles}, where Eccosorb is typically used to reduce quasiparticle-based losses caused by stray IR photons\cite{barends2011minimizing,EccosorbCorcoles}. We examined whether quasiparticles generated from IR photons are important for nanowire-embedded resonators by measuring them in an Eccosorb-lined sample box. As the hollow dotted symbols in Fig.~\ref{fig3}c show, losses at high $\left<n\right>$ are much lower than for the standard sample box and $Q_{\rm i} \approx$~6--9$\times$10$^5$. This value matches that of the bare resonators for the same $\left<n\right>$, suggesting that the density of residual quasiparticles has been reduced to that of the bare resonators (see Table~\ref{ResTab1}). A saturated high-$\left<n\right>$ $Q_{\rm i}$ is not observed, due to nonlinearities in the resonance lineshape of the nanowire-embedded resonators. With the smaller quasiparticle-based loss, the TLS-based low-$\left<n\right>$ trend of loss increasing as $\left<n\right>$ decreases is once again found. The high-$\left<n\right>$ $Q_{\rm i}$ is found to increase with increasing temperature, consistent with losses from thermally generated quasiparticles as found in the bare resonators, indicating that increased quasiparticle losses in nanowire-embedded resonators in the normal sample box arose from quasiparticles excited by IR photons. As Table~\ref{ResTab1} shows, $\delta_{qp}$ of the nanowire-embedded resonators in the Eccosorb environment match those of the bare resonators (both with and without the Eccosorb environment) and are therefore limited by another mechanism which is not unique to the nanowire.

%Figure~\ref{fig3}b shows a measurement of loss for the same resonator with and without the Eccosorb enclosure. The high-$\left<n\right>$ loss is found to decrease only slightly when the Eccosorb-lined sample box is used, in contrast to nanowire-embedded resonators. In fact, it is unsurprising that the Eccosorb has a small effect in the bare resonators since the energy gap of NbN is $\approx\times$10 larger than in Al, but the reason for the sensitivity to IR photons in the nanowire-embedded resonators is not immediately obvious.

Figure 3b shows the loss for the same bare resonator with and without the Eccosorb enclosure. In contrast to nanowire-embedded resonators, the high-$\left<n\right>$ loss decreases only slightly when the Eccosorb-lined sample box is used. This is actually unsurprising since the energy gap of NbN is $\sim{10}\times$ larger than in Al. On the other hand, the reason for the sensitivity to IR photons in the nanowire-embedded resonators is not immediately obvious.
 %It is {\it a priori} possible that there is a suppression of the superconductivity in the nanowire. To explore this, we determine the critical current $I_{\rm c}$ from measurements of the resonator's bifurcation\cite{jenkins2014nanoscale}. Fig.~\ref{Sfig2}a \& b (c \& d) show the bare (nanowire-embedded) resonator's transmission response at 307~mK as a function of the applied microwave power. The profile lines show respectively the low-power response of the resonator (blue), an asymmetric resonator response at increased transmission power (green), and the bifurcation in the resonator response at high transmission powers (red). Since $W_{\rm sto} = \left<LI^2\right>$, where $L$ is the total inductance of the resonator and $I$ is the amplitude of the microwave current at the current antinode, the bifurcation power enables $I_{\rm c}$ to be determined. To allow comparison of bare and nanowire-embedded resonators, in table~\ref{ResTab1} we express $I_{\rm c}$ as a critical current density, $J_{\rm c}$. For both bare and nanowire-embedded resonators, with and without Eccosorb $J_{\rm c}$ is of order 10$^9$~A/m$^2$. This indicates that the superconductivity of the nanowire-embedded resonators has not been significantly suppressed.
Our results demonstrate the importance of IR filtering even when nanowires have a large superconducting energy gap such as those in NbN. This is relevant to all nanowire-based devices. We note that a small suppression of $T_{\rm c}$ in our nanowire (below the precision of our $T_{\rm c}$ determination) could give some enhanced sensitivity to IR photons. Alternative explanations for the sensitivity include the nanowire exhibiting a different quasiparticle lifetime\cite{de2011number} or non-equilibrium superconductivity\cite{goldie2012non}, but these are beyond the scope of this study, although, since $Q_{\rm i}$ remains high, the number of quasiparticles created from IR photons must still be quite small\cite{barends2011minimizing}.

Finally, we compare the consistency of the TLS-loss rates (Table~\ref{ResTab1} and supplemental\cite{suppnotePRApp}) obtained from the analysis of the data shown in Figs.~\ref{fig4}~\&~\ref{fig3}. $F\delta_{\rm TLS}^{i}$ and $F\delta_{\rm TLS}^{0}$ differ by less than 20\%, this difference is because $F\delta_{\rm TLS}^{0}$ is only sensitive to near-resonant TLS, whereas $F\delta_{\rm TLS}^{\rm i}$ is also sensitive to a broad spectrum of off-resonant TLS\cite{macha2010losses,bruno2015reducing,pappas2011two}. Next, we note that $\delta^{i}_{\rm TLS} = \chi$,\cite{faoro2015interacting} so that the ratio $FP_{\gamma}\chi$/$F\delta^{i}_{\rm TLS}$ gives $P_{\gamma}$. We find an average value of $P_{\gamma} =$~0.093. This agrees well (see supplemental\cite{suppnotePRApp}) with the charge noise spectra of single-electron transistors that give $P_{\gamma} \approx $~0.10.\cite{kafanov2008charge,ratesnote} Therefore, all TLS-loss rates are consistent with each other. The TLS loss rates imply a TLS-limited $Q_{\rm i}$ up to $\approx$2$\times$10$^5$ in the quantum limit (at temperatures down to 10~mK and at single-photon energies). This is approximately 100$\times$ larger than in equivalent nanowire-embedded resonators and compares favourably with Josephson-junction-embedded resonators.

\section{Conclusion}

To conclude, we have used a neon FIB to create superconducting nanowires with widths down to 20~nm within superconducting resonators. In the low-power limit, these devices demonstrated $Q_{\rm i}$ up to 3.9$\times$10$^5$ at 300~mK, with $\delta^{\rm i}_{\rm TLS}$ and $\delta^{0}_{\rm TLS}$ corresponding to a TLS-limited $Q_{\rm i}$ up to 2$\times$10$^5$ at 10~mK. These TLS losses arise from the NbN thin-film technology rather than the neon FIB, meaning a higher $Q_{\rm i}$ should be possible with better resonator technology\cite{bruno2015reducing}. By obtaining such a high $Q_{\rm i}$ using nanowires, we have demonstrated a critical step towards realising nanowire-based, superinductance,  phase-slip or Dayem-bridge circuits with coherence times comparable to conventional Josephson-junction-type devices. 

\section{acknowledgements}The authors would like to acknowledge useful discussions with S.~de Graaf, E.~Dupont-Ferrier, N.~Constantino and T.~Lindstr\"{o}m. We also thank T.~Lindstr\"{o}m for the loan of equipment and critical reading of the manuscript. The authors gratefully acknowledge funding from the UK EPSRC, grant references EP/J017329/1 (JB and JCF) and EP/K024701/1 (JS and PAW), and Carl Zeiss SMT (JS and PAW). \nocite{santavicca2008impedance,faoro2015interacting,goetz2016loss,barends2011minimizing,gustavsson2016suppressing,budoyo2016effects,macha2010losses,burnett2016analysis,wisbey2010effect,khalil2011loss,faoro2012internal,lisenfeld2015observation,skacel2015probing,burnett2014evidence,burnett2016analysis,ramanayaka2015evidence,kafanov2008charge,mattis1958theory,semenov2009optical,coumou2015electrodynamics,de2014evidence,adamyan2016tunable}

\bibliography{refs}

\begin{thebibliography}{47}
\bibitem [{rat()}]{ratesnote}%
  \BibitemOpen
  \href@noop {} {}\bibinfo {note} {This arises from $P_{\gamma} =
  1/\ln(\gamma_{\rm max}/\gamma_{\rm min})$, where $\gamma_{\rm max}$
  ($\gamma_{\rm min}$) is the maximum (minimum) rate of TLS switching. For a
  TLS charge noise spectrum extending from a few Hz to a few 100kHz,
  $P_{\gamma} \approx$~0.1.}\BibitemShut {Stop}%
\end{thebibliography}

%\begin{figure} 
%\hspace*{-2cm}\includegraphics{NWsup-resub-a.pdf}
%\end{figure}

\end{document}